\documentclass[aps,prb,twocolumn,floatfix,footinbib,showpacs]{revtex4}

\usepackage{epsfig}
\usepackage{amsfonts}
\usepackage{amsmath}
\usepackage{amssymb}
\usepackage{graphicx}
\usepackage{subfigure}
\begin{document}

\preprint{\today}

\title{Mapping between quantum dot and quantum well lasers: 
From conventional to spin lasers}

\author{Jeongsu Lee}
\email{jl376@buffalo.edu}

\author{Rafa{\l} Oszwa{\l}dowski}
\email{rmo4@buffalo.edu}

\author{Christian G{\o}thgen}

\author{Igor \v{Z}uti\'{c}}
\email{zigor@buffalo.edu}

\affiliation{Department of Physics, University at Buffalo--SUNY, 
Buffalo, NY 14260, USA 
}

\date{\today}

\begin{abstract}
We explore similarities between the quantum wells  and quantum dots used as 
optical gain media
in semiconductor lasers. We formulate a mapping procedure 
which allows a simpler, often analytical, description of quantum well lasers 
to  study more complex lasers based on quantum dots. The key observation
in relating the two classes of laser is that the influence of a finite capture time on 
the operation of quantum dot lasers can be approximated well by a suitable choice
of the gain compression factor in quantum well lasers. Our findings are applied
to the rate equations for both conventional (spin-unpolarized) and spin lasers 
in which spin-polarized carriers are injected optically or electrically. 
We distinguish two types of mapping that pertain to the steady-state and dynamical 
operation respectively and elucidate their limitations.
\end{abstract}


\maketitle

\section{Introduction}
\label{intro}
The importance of lasers typically reflects two aspects: their practical use 
in a wide range of applications and their highly controllable 
nonlinear coherent optical response.\cite{Chuang:2009,Parker:2004,Coldren:1995,
Chow:1999,Haken:1985}  In addressing the first aspect there is 
a systematic effort to reduce the required injection for the onset of lasing. 
In semiconductor lasers, this can be realized by fabricating structures of 
reduced dimensionality, such as quantum wells, wires, and 
dots,\cite{Alferov2001:RMP,Ustinov:2003,Bimberg:1999}
or by considering physical mechanisms that enhance stimulated emission,
such as polaritons or introduction of spin-polarized 
carriers.\cite{Das2011:PRL, Rudolph2003:APL,Rudolph2005:APL,Holub2007:PRL}
In the second aspect, lasers also present valuable model systems to elucidate 
connections to other cooperative phenomena.\cite{Haken:1985,Degiorgio1970:PRA}
As the injection or pumping of the lasers is increased, there is a transition from incoherent 
to coherent emitted light that can be described by
the Landau theory of second-order 
phase transitions.\cite{Haken:1985}
Moreover, the instabilities found in lasers 
directly resemble instabilities found in electronic devices.\cite{Degiorgio1976:PT}
Since some lasers provide highly accurate and tunable parameters, further insights 
can be achieved by establishing mapping procedures between such lasers and 
other cooperative phenomena, such as ferromagnetism.
\cite{Haken:1985,Degiorgio1970:PRA, Degiorgio1976:PT}

In this work we explore similarities between the quantum wells (QWs) 
and quantum dots (QDs) used as the gain material 
in semiconductor lasers. On one hand, QW lasers have a very transparent description, 
readily available at the textbook level.\cite{Chuang:2009,Parker:2004}
On the other hand, while QD-based active regions 
require a more complicated description, they also lead to desirable 
operation properties, such as low threshold for lasing, robust temperature 
performance, low chirp, and narrow gain spectra.\cite{Asryan:2002, Sellers2004:PRB}
Therefore a mapping between QD- and QW-based lasers
has the potential to yield a simple description 
(as used in QW lasers) to investigate a more complex and yet technologically 
interesting systems (involving QDs).

To establish such a mapping
we focus on two cases: (i) conventional (spin-unpolarized) lasers, and (ii) 
spin lasers in which the spin-polarized carriers are injected by circularly 
polarized light or by electrical injection (using a magnetic contact). 
Spin lasers can be described as a generalization of conventional
lasers: with spin-unpolarized injection, spin lasers must reduce 
to conventional lasers.\cite{Rudolph2003:APL, Gothgen2008:APL, 
Lee2010:APL,Oszwaldowski2010:PRB}
A further motivation to consider spin lasers in the current context 
is provided by the recent 
experiments showing significant improvements in 
QD-lasers\cite{Basu2008:APL,Basu2009:PRL,Saha2010:PRB} 
(including 100 K  higher operation than in their electrically 
injected QW-based counterparts\cite{Holub2007:PRL}), 
which were analyzed as if they were QW lasers. 

\begin{figure}[htbp]
\subfigure[]{
\includegraphics[scale=0.8]{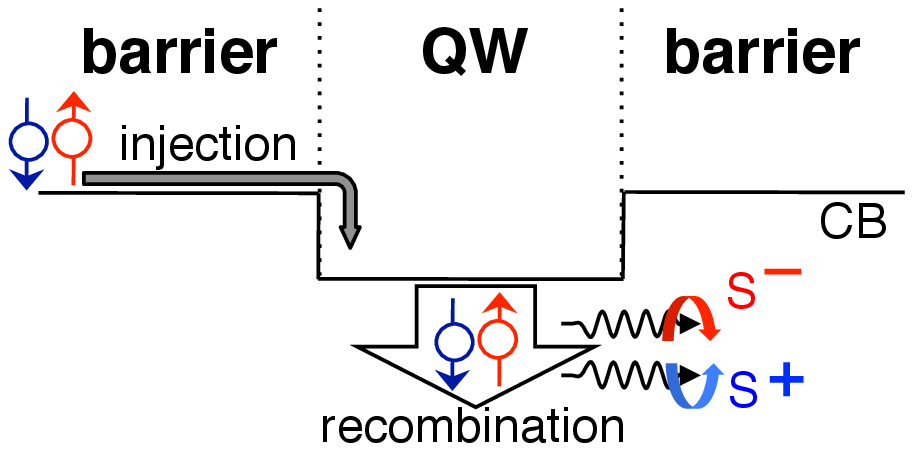}
\label{fig:01a}
}
\subfigure[]{
\includegraphics[scale=0.8]{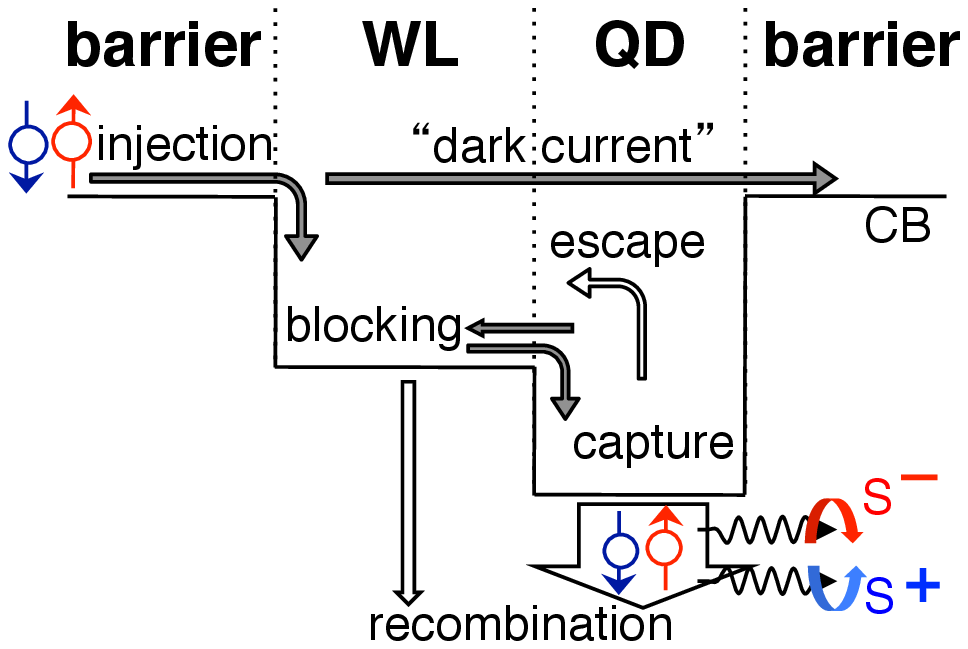}
\label{fig:01b}
}
\caption{(Color online) Conduction band (CB) diagram and characteristic processes in
semiconductor lasers. (a)  Quantum well (QW) laser. A preferential
spin alignment of the injected carriers, leads, 
through electron-hole recombination, to circularly polarized emitted light  
($S^\pm$ are the emitted
photons with positive and negative helicity, respectively). 
(b)  Quantum dot (QD) laser contains an additional level arising from the 
wetting layer (WL) as well as several more processes, not present in QW lasers.
\label{fig:01}}
\end{figure}

A schematic description of the QW and QD semiconductor laser is depicted
in Fig.~\ref{fig:01}, representing the conduction band (CB) diagram 
and several characteristic processes included in our rate equation (RE) 
approach.\cite{Gothgen2008:APL,Lee2010:APL,Oszwaldowski2010:PRB}
With the usually employed assumption of charge neutrality, 
the underlying picture is simplified since holes need not be explicitly 
considered for conventional lasers.\cite{Gothgen2008:APL} 
The injection of spin-polarized carriers leads to circular 
polarization of the emitted light.\cite{Meier:1984,Zutic2004:RMP,%
Zutic2001:PRB,Vulovic2011:JAP,Iba2011:APL} 
Depicted carrier recombination (in both QWs and QDs) is either spontaneous
or stimulated, and a sufficiently high injection leads to the onset of lasing
when the optical gain can overcome losses in the resonant cavity.

A more complex description of QD lasers 
includes several additional processes and a two-dimensional QW-like 
wetting layer (WL), which acts as a reservoir of carriers.\cite{Dery2005:JQE,%
Fiore2007:JQE,Summers2007:JAP}  Carriers
from the WL are captured to the QD or, conversely, they can escape from
QD to WL. To correctly describe the small density of QD states, as well
as saturation of the WL states at high injection, it is important to include 
the Pauli blocking\cite{Oszwaldowski2010:PRB, Dery2005:JQE,Fiore2007:JQE,%
Summers2007:JAP,Taylor2010:APL}
which impedes carrier transfer to states close to saturation.
The Pauli blocking is responsible for additional nonlinear contributions 
to the QD REs and for a dark current (i.e., a current that is not  
accompanied by any emission of light), both of which are absent 
in our simpler description of QW lasers.

\begin{figure}[htbp]
 \subfigure[]{
   \includegraphics[scale =0.28] {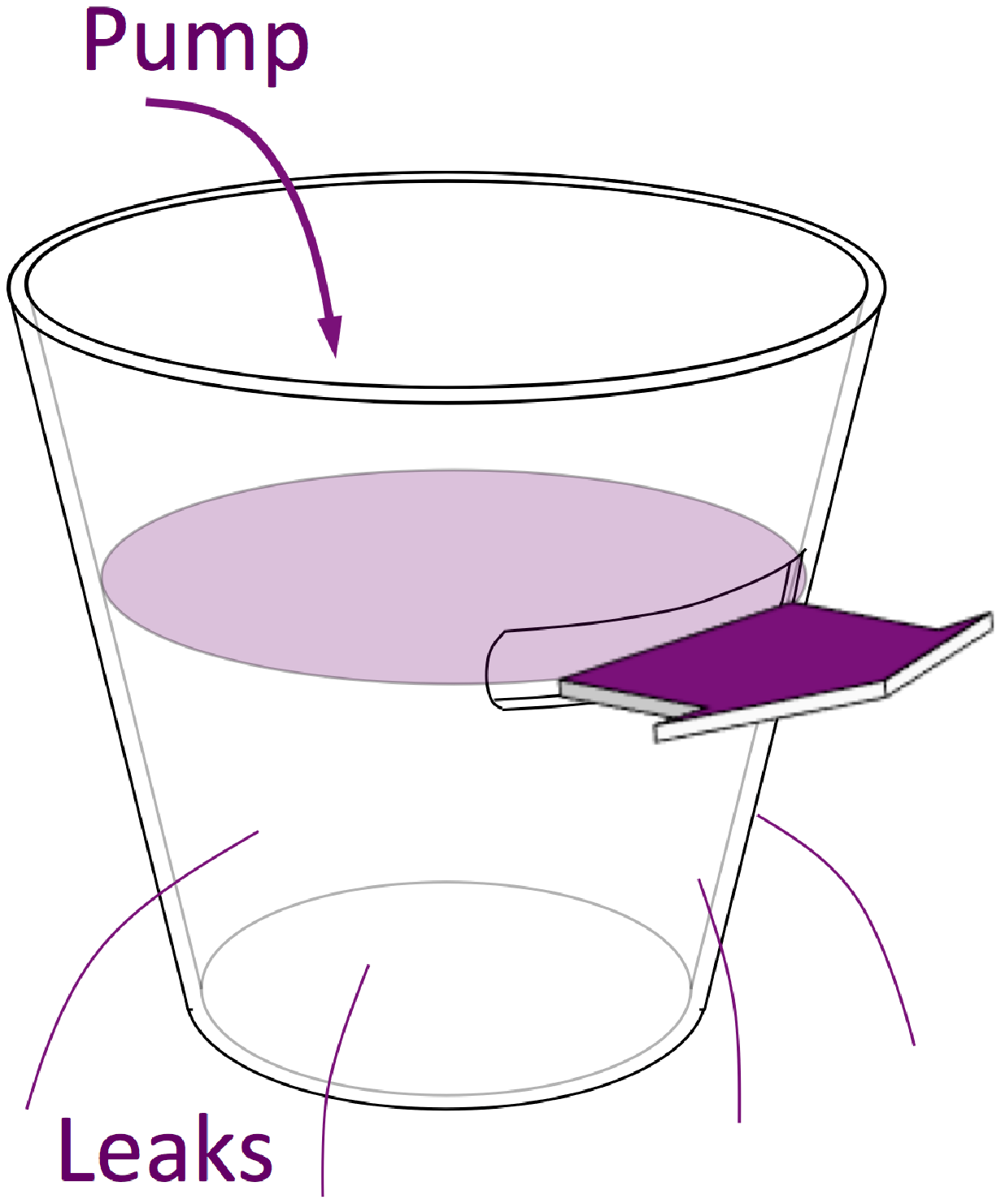}
\label{fig:02a}
}
\subfigure[]{
   \includegraphics[scale =0.28] {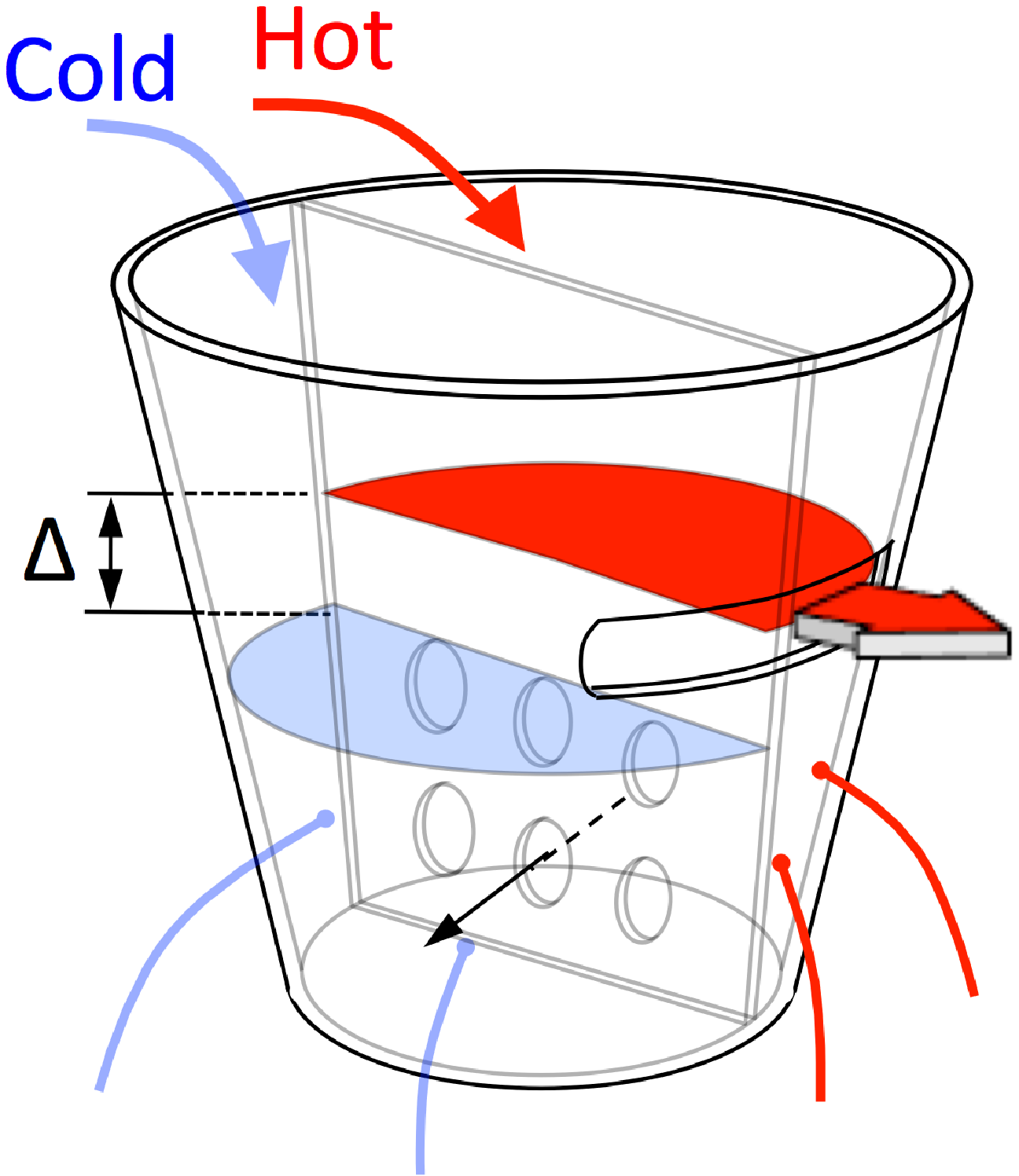}
\label{fig:02b}
}

\caption{(Color online) 
Bucket model of lasers. (a) Conventional laser. Pump (injection)
fills the bucket with small leaks (corresponding to spontaneous recombination
and the {\em off} regime with negligible light emission) and a large slit from which,
above sufficiently strong pumping, the water will gush out (corresponding to 
stimulated emission and the onset of lasing in the {\em on} regime). An additional
increase in pumping will lead only to a small change in the water level (representing
carrier density in a laser), but the output will increase rapidly, as compared to the
{\em off}  regime. (b) Spin laser. Two halves of the bucket, representing two spin
populations (hot and cold water) are separately filled. The partition between them
is not perfect: openings in the partition model the spin relaxation which mixes the 
two populations. The difference between uneven water levels,
denoted by $\Delta$, represents
the spin imbalance in the laser. Here, in addition to the {\em on} and {\em off} regimes,
one can infer a regime where only hot water will gush out. This
represents the spin-filtering regime between two different lasing thresholds: even a 
modest polarization of injection leads to complete polarization of emission.} 
\label{fig:02}
\end{figure}

To provide an intuitive picture of changes arising from the spin-polarized
injection, we develop here a bucket model of spin lasers,  compared 
in Fig.~\ref{fig:02} with the well-known model for conventional 
lasers.\cite{Parker:2004}  A simple analogy with the pumped bucket illustrates
{\em on} and {\em off} regimes in conventional lasers, where the outgoing
water represents the emitted light. At low injection or pumping $J$, there is only
negligible output light. The operation of a laser is similar to that of a light emitting 
diode (LED); the spontaneous recombination is responsible for the emitted light. 
At higher injection, when the water starts to gush out of the large slit 
in Fig.~\ref{fig:02a}, the lasing threshold is reached. At the threshold 
injection $J_T$, stimulated emission starts and the emitted light intensity  
increases significantly. $J>J_T$ corresponds to lasing operation in which 
the stimulated recombination is the dominant mechanism of light emission.

We next turn to the pictorial representation of a simple spin laser.
To model different projections of carrier spin or helicities of light, it
is convenient to think of an analogy with hot and cold water, 
as shown in Fig.~\ref{fig:02b}.
The bucket is partitioned into two halves, representing two spin populations,
which are separately filled with hot and cold water, respectively. The openings in 
the partition allow mixing of hot and cold water, intended 
to model the spin relaxation.\cite{Zutic2004:RMP} With an unequal injection of hot 
and cold water, injection spin polarization is defined as\cite{negative} 
\begin{equation}
P_J= (J_+ - J_-)/J,
\label{eq:pj}
\end{equation}
where $J_\pm$ are the injections of the two spin projections which together 
comprise the total injection  $J= J_+ + J_-$. The difference in the hot and 
cold water levels $\Delta$ [see Fig.~\ref{fig:02b}], leads to the three operating 
regimes and two different lasing thresholds $J_{T1,2}$.\cite{Gothgen2008:APL} 

At low pumping (when both hot and cold water levels are below the large slit), 
both spin-up and spin-down carriers are in the {\em off} (LED) regime,  
thus with negligible emission. At higher pumping, the hot water 
reaches the large slit and it gushes out as depicted in Fig.~\ref{fig:02b}, while the amount
of cold water coming out is still negligible. 
Such a scenario represents
a regime in which the majority spin is lasing, while the minority spin 
is still in the LED regime; thus the stimulated emission is from recombination of 
majority spin carriers. 
Two important consequences of this regime are already 
confirmed experimentally:
(i) A spin laser will start to lase at a smaller total injection than a corresponding 
conventional laser (only a part of the bucket needs to be filled). This represents 
the threshold reduction in 
spin lasers,\cite{Rudolph2003:APL,Holub2007:PRL,Gothgen2008:APL,%
Vurgaftman2008:APL,Oestreich2005:SM,Holub2011:PRB} which can be parametrized as
\begin{equation}
r= 1-J_{T1}/J_T,
\label{eq:r}
\end{equation}
where $J_{T1}$ is the majority spin threshold ($J_{T1}<J_T$).
(ii) Even a modest injection polarization $P_J \ll 1$ can lead to highly 
circularly polarized light.\cite{Lee2010:APL,Saha2010:PRB} 
The relative width of this ``spin-filtering regime'' can be expressed as 
the interval\cite{Oszwaldowski2010:PRB}
\begin{equation}
d=(J_{T2}-J_{T1})/J_T,
\label{eq:d}
\end{equation}
where $J_{T2}$ is the minority spin threshold 
($J_{T1}< J_T< J_{T2}$) and the width of this interval
increases with the injection spin polarization.
$J>J_{T2}$ gives rise to minority helicity photons from minority spin carriers,
and the spin polarization of light converges to $-P_J$ with increasing 
injection,\cite{Gothgen2008:APL,negative} 
analogous to the situation where both hot and cold water gush out. 

Based on the intuitive pictures that we describe here, 
we present the rate equations for QD and QW  lasers in the following section.
Solution of the REs in the steady-state and dynamic-operation regimes gives
two possible approaches to the mapping.
In Sec.~\ref{SSM}, we focus on the steady-state mapping 
while Sec.~\ref{DOM} is dedicated to the dynamic mapping of conventional lasers.
We analyze the differences between these two mapping methods
in Sec.~\ref{SSM vs DOM} because the two mappings are not equivalent to each other.
In Secs.~\ref{MOS} and \ref{SSA}, we expand the mappings to spin lasers. 
Finally, we summarize our work; and suggest possible directions for further research.

\section{Rate Equations}
\label{REs}

In this work  we consider rate equations which have been successfully used  
to describe both conventional and spin lasers.\cite{Chuang:2009, Parker:2004,%
Rudolph2003:APL,Rudolph2005:APL,Holub2007:PRL,Gothgen2008:APL,%
Basu2008:APL,Basu2009:PRL,Saha2010:PRB, Fiore2007:JQE,%
Summers2007:JAP,SanMiguel1995:PRA,Dyson2003:JOB,Adams2009:JQE,%
Al-Seyab2011:PJ,Gerhardt2011:APL, Jahme2010:APL,Dery2004:JQE}
An advantage of this approach is its simplicity. REs can provide a direct 
relation between material characteristics and device 
parameters,\cite{Fiore2007:JQE} as well as often allowing analytical solutions and 
an  effective method to elucidate many trends in the operation of 
lasers.\cite{Chuang:2009,Parker:2004,Coldren:1995,Gothgen2008:APL,Lee2010:APL} 
For conventional QW lasers, we employ the widely used REs (Ref.~\onlinecite{Chuang:2009}) 
for carrier and photon density, $n$ and $S$, respectively (generalized REs for 
spin lasers are given in Appendix A):
\begin{eqnarray}
dn/dt&=&J -g(n,S)S -R_\mathrm{sp}, \label{eq:REn} \\  
dS/dt&=&\Gamma g(n,S)S +\Gamma \beta R_\mathrm{sp} -S/\tau_\mathrm{ph}, \label{eq:RES}
\end{eqnarray}
where the charge neutrality was used to eliminate the REs for holes. 
To describe stimulated emission, the optical gain term is usually modeled 
as\cite{Chuang:2009} 
\begin{equation}
g(n,S)=g_0(n-n_{\mathrm{tran}})/(1+\epsilon S), \label{eq:gain}
\end{equation}
where $g_0$ is the gain coefficient,\cite{Holub2007:PRL} 
$n_{\mathrm{tran}}$ is the transparency density at which 
the optical gain becomes zero,
and $\epsilon$ is the gain compression factor.\cite{Chuang:2009,Lee1993:OQE}
The spontaneous recombination $R_\mathrm{sp}$ can have various density dependences;
here we focus on the quadratic form\cite{Chuang:2009,Gothgen2008:APL,Zutic2006:PRL} 
$Bn^2$, where $B$ is a temperature-dependent constant.
$\Gamma$ is the optical confinement factor, arising from different volumes of the resonant 
cavity and the active region of the lasers;\cite{Chuang:2009} 
$\beta$ is the spontaneous emission factor ($\beta \rightarrow 0$
is an accurate approximation since typical experimental values 
$\beta \sim 10^{-5}-10^{-4}$ do not alter laser behavior 
significantly,  but slightly complicate the definition of threshold).%
\cite{Gothgen2008:APL,Lee2010:APL,Yu:2003}
The photon lifetime $\tau_\mathrm{ph}$ reflects optical losses
such as absorption in the boundary media, photon scattering, and loss at the mirrors.\cite{Yariv:1997}

To describe QD lasers, it is more appropriate to use occupancies,
rather than carrier and photon densities.\cite{Fiore2007:JQE,%
Summers2007:JAP} The REs describing the QD-based lasers 
[Fig.~\ref{fig:01b}] are more complex than  Eqs.~(\ref{eq:REn}) and 
(\ref{eq:RES}), used for QW lasers. REs for QD spin lasers 
are given in Appendix A. Here we explain their limiting case for
conventional lasers written in an abbreviated form,
\begin{eqnarray}
df_{w}/dt&=&I-C+\frac{2}{\kappa}E-R_{w} \label{eq:fw},\\
df_{q}/dt&=&\frac{\kappa}{2}C-E-R_{q}-G \label{eq:fq},\\
df_{S}/dt&=& \Gamma_\mathrm{QD} G+ \Gamma_\mathrm{QD} \beta R_q- f_{S}/\tau_\mathrm{ph}, \label{eq:fS} 
\end{eqnarray}
where the indices $w$ and $q$ represent the WL and QD regions,
while the index $S$ pertains to photons. The electron occupancies
(those for holes were eliminated using charge neutrality and the assumption 
that the capture and escape times for the electrons and holes are equal)  
$0\le f_{w,q} \le 1$  are related to the corresponding number of electrons
$\bar{n}_{w,q},$ as $f_w=\bar{n}_w/N_w$ and $f_q=\bar{n}_q/(2N_q)$, 
where $N_w$ is the number of states in the WL and $N_q$ is the number of QDs
[each dot contains a twofold- (spin-) degenerate level], related by the
ratio $\kappa=N_w/N_q$.
Here, we use an overbar to distinguish numbers 
from the corresponding densities used in Eqs.~(\ref{eq:REn})-(\ref{eq:gain}). 
The photon occupancy 
$f_S=\bar{S}/(2N_q)$,  where $\bar{S}$ is the number of cavity photons, does 
not have an upper bound. 

The carrier injection and the capture from the WL to the QDs are 
$I=j(1-f_w)$ and $C=f_w(1-f_q)/\tau_c$, where $j$ is the number of
carriers (electrons) injected into the laser per WL state and unit time,
while $\tau_c$ is the capture time.  An opposite process to the carrier 
capture is their escape $E=f_q(1-f_w)/\tau_e$, where $\tau_e$
is the escape time. These processes have a characteristic Pauli  
blocking factor $(1-f)$, absent in the analysis of QW lasers, 
as shown in Eqs. (\ref{eq:REn}) and  (\ref{eq:RES}). It is instructive to note
the nonlinear form (in the carrier occupancies) of the escape and 
capture terms $E$, $C$ in QDs. 
The absence of such nonlinearities in QW laser REs provides another simplification in 
understanding QD lasers through the mapping procedure, which allows 
their more transparent description.
Other processes depicted in Fig.~\ref{fig:01b} are the
spontaneous radiative recombinations 
$R_\eta=b_\eta f^2_\eta$, where $\eta=w,q$. 
The charge neutrality implies that $f^2_\eta$ actually corresponds
to the product of electron and hole occupancies.\cite{Oszwaldowski2010:PRB} 
Coupling of carriers and light in Eqs.~(\ref{eq:fq}) and (\ref{eq:fS}) 
is responsible for  stimulated emission, which can be described by
\begin{equation}
G=g(2 f_q-1) f_S,
\label{eq:G} 
\end{equation}
where $g$ is independent of photon occupancies and does not
contain the gain compression factor $\epsilon$, used in the QW lasers.
By using occupancies, rather than densities,  for QD REs, 
different volume factors are eliminated and there
is no need to introduce the optical confinement factor ( $\Gamma_\mathrm{QD}=1$), 
required in 
Eq.~(\ref{eq:RES}). Finally, $\tau_\mathrm{ph}$ is analogous to
the quantity  already used in Eq.~(\ref{eq:RES}). 

\section{Steady-State Mapping}
\label{SSM}

Based on the REs described in Sec.~\ref{REs}, we explore the feasibility 
of mapping between QD and QW lasers.  Our goal is to 
approximate the solutions for the more complicated QD laser REs by the 
solutions we obtain from solving REs for QW lasers. To achieve the mapping,
there are two requirements for the mapped QW laser REs. First, the REs should be able
to estimate steady-state properties such as threshold and light intensity
with a reasonable accuracy. At the same time, the dynamic response of
lasers should also be presented by the REs through various numerical or
analytical methods such as large- or small-signal analyses.
In this section, as the first step, we focus on 
the steady-state operation for which the QW mapping parameters  are
extracted from the QD parameters by 
solving Eqs.~(\ref{eq:REn}), (\ref{eq:RES}) and (\ref{eq:fw})-(\ref{eq:fS}) analytically,
while the constants $\tau_\mathrm{ph}$ and $\beta$ are kept the 
same for QD and QW lasers.
This mapping corresponds to the situation in which the
active region comprised of QDs and WLs is considered as QWs,
while retaining the remaining geometry of the laser.
Ideally, the following equations should hold:
\begin{eqnarray}
J&=& \kappa (N_q/V) j,\label{eq:Jmap}\\
n(J)&=& (N_q/V) [2f_q(j)+\kappa f_w(j)],\label{eq:nmap}\\
S(J)&=&2 \Gamma (N_q/V) f_S(j),\label{eq:Smap}
\end{eqnarray}
where $N_q$ and  $\kappa$ were previously defined,  
$V$ is the volume of the active region,
and $J$ ($j$) represents injection in the QW (QD) laser REs. 
While Eq.~(\ref{eq:Jmap}) holds by definition, Eqs.~(\ref{eq:nmap}) and (\ref{eq:Smap}) 
can not be satisfied for all $j$'s.
Therefore we impose four matching points where 
the two solutions from QW and QD laser REs  should coincide:

(i) Transparency carrier density 
$n_\mathrm{tran}= (N_q/V) [2f_q(j_{\mathrm{tran}})+\kappa f_w(j_{\mathrm{tran}})]$,
where $n_\mathrm{tran}=n(J_\mathrm{tran})$, and 
$J_\mathrm{tran}$ and $j_{\mathrm{tran}}$ are the injection values at transparency 
satisfying $f_q(j_{\mathrm{tran}})=1/2$ [see Eqs.~(\ref{eq:gain}) and (\ref{eq:G})].

(ii) Threshold carrier density 
$n_{T}= n_{\mathrm{tran}}+1/ (\Gamma g_0 \tau_\mathrm{ph})
= (N_q/V) [2f_q(j_T)+\kappa f_w(j_T)]$ 
determines $g_0$ (for QW lasers), 
where $j_T$ is the injection (for QD lasers) at threshold. 

(iii) Threshold current density
$J_T=B n_T^2= \kappa (N_q/V) j_T$ determines $B$ (for QW lasers), 
where $j_T$ is the QD laser threshold injection current. 

(iv) Photon density
$S=2 \Gamma (N_q/V) f_S$ determines $\epsilon_s$, 
where the subscript $s$ denotes that it was obtained
in the steady-state (static) case. At a fixed injection ($J=\alpha J_T$ or $j=\alpha j_T$),
the photon densities obtained from QD and QW laser REs are made to coincide. 
In this paper, $\alpha=10$ is used.

\begin{figure}
\includegraphics[width=\columnwidth]{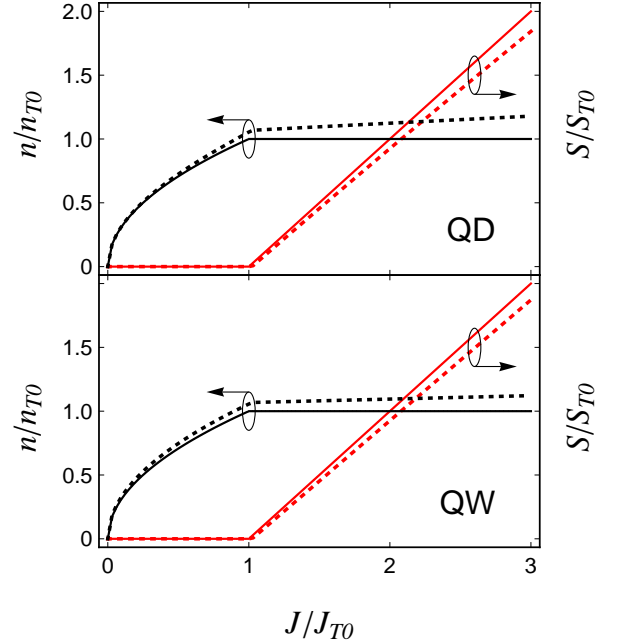} 
\caption{(Color online) 
Carrier ($n$) and photon densities ($S$) as functions of injection $J$.
$n$ and $J$ are normalized to their threshold values $n_{T0}$ and $J_{T0}$ 
for capture time $\tau_c=0$ (QD lasers) or gain compression factor $\epsilon_s=0$ (QW lasers),
while $S$ is normalized to $S_{T0}\equiv S(2J_{T0})$.
For QD lasers, solid and broken lines represent $\tau_c=0$ and $\tau_c=2$ ps,
while they represent $\epsilon_s=0$ and $1.62\times10^{-14}~$cm$^3$ for QW lasers.
Note that QW laser characteristics for $\epsilon_s=0$ are identical to QD laser characteristics for 
$\tau_c=0$. 
The mapping parameters used throughout this paper are given in Table \ref{tab:01}.
\label{fig:03}}
\end{figure}

\begin{table}
\caption{Mapping parameters\label{tab:01}. 
The QD parameters are $\tau_\mathrm{ph}=2$ ps,
$b_q \tau_\mathrm{ph}$=0.01, $b_w \tau_\mathrm{ph}$=2.33, 
$g \tau_\mathrm{ph}=2$, $\kappa$=100 and $\tau_e$=1 ns (Ref.~
\onlinecite{Fiore2007:JQE,Ezra2009:JQE,Erneus2007:PRA,Rsp:comment}).}
\setlength{\tabcolsep}{10pt}
\begin{tabular}{llll}
\hline \hline
QW param. & $\tau_c=0$ & $\tau_c=2$ ps & unit\\
\hline
$\epsilon_{s}$ & 0 & $1.62\times10^{-14}~$ & cm$^3$\\
$\epsilon_{d}$ (Ref.~\onlinecite{sec:DOM}) & 0 & $6.39\times10^{-15}~$ & cm$^3$\\
$g_0$ & 1.90 $\times10^{-3}$ & 1.65 $\times 10^{-3}$ & cm$^{3}$s$^{-1}$\\
$n_{\mathrm{tran}}$ & $3.50\times10^{16}$ & $3.58\times10^{16}$ & cm$^{-3}$ \\
$B$ & $1.43\times10^{-7}$ & $1.28\times10^{-7}$ & cm$^{3}$s$^{-1}$\\
$\tau_\mathrm{ph}$ & \multicolumn{2}{c}{2} & ps \\
$\Gamma$ & \multicolumn{2}{c}{0.03} & \\
$\beta$ & \multicolumn{2}{c}{0} & \\
\hline \hline
\end{tabular}
\end{table}

Results of the above mapping are shown in the Fig.~\ref{fig:03},
comparing the light-injection and carrier-density-injection
characteristics for QD and QW lasers for different $\tau_c$'s. 
The corresponding mapping parameters are given in Table~\ref{tab:01}.
In the limit of $\tau_c\to0$, the wetting layer is ``transparent'' to 
carriers, since injected carriers are immediately captured into QDs; 
therefore, within the RE description for $\epsilon_s=0$,  
QD and QW lasers behave identically.
In both cases, when  $\tau_c=0$ or $\epsilon_s=0$, the carrier density 
(black solid) is pinned (fixed) above the threshold and the photon
density increases linearly, as expected.\cite{Chuang:2009,Saha2010:PRB}

For finite $\tau_c$ (2 ps in Fig.~\ref{fig:03}) 
the carrier density is enhanced and the photon density is
suppressed. For $n> n_T$ the increase in carrier density 
can be mostly attributed to WL occupancy, which increases for $\tau_c>0$.
Without any gain term in Eq.~(\ref{eq:fw}), we can infer that the 
WL occupancy does not contribute to stimulated emission. However,
since the active region is considered as comprised of QDs and the WL, we
take into account this ``inefficiency'' of the WL for light emission by 
introducing $\epsilon_s$ in the QW laser model. 
The typical range of $\epsilon$ experimentally 
obtained in QW lasers\cite{Holub2007:PRL,Rudolph2003:APL} 
is $\epsilon \sim 10^{-19}-10^{-17}$. In contrast,  in our mapping we  
employ  (Table~\ref{tab:01}) a much larger $\epsilon_s\sim10^{-14}$, which
captures well the behavior of QD lasers, as can be seen by comparison
of the upper and lower panels in Fig.~\ref{fig:03}.
The analysis of Fig.~\ref{fig:03} reveals that in a QD laser the effect 
of finite $\tau_c$ (i.e.,~to increase $n$ and suppress $S$ with injection) is 
similar to the influence of a finite gain compression factor $\epsilon_s$ 
in a QW laser. This suggests that, in the steady state, 
by finding $\epsilon_s$ as a function of $\tau_c$,
QD laser REs can be accurately replaced by QW laser REs.
It is also instructive to note that as $\tau_c$ becomes longer in QD lasers, 
parasitic effects such as spectral hole burning or phonon bottleneck 
will arise, and these nonlinear effects are taken into account by introducing $\epsilon$ 
in QW laser REs.

\section{Dynamic-Operation Mapping}
\label{DOM}

The mapping between QD and QW lasers works well in the steady state,
and the influence of finite $\tau_c$, as shown in 
Fig.~\ref{fig:03}, on light-injection and 
even carrier-density-injection characteristics is accurately modeled 
by introducing a large $\epsilon_s$ in the QW laser. However, the most useful
properties of lasers typically pertain to their dynamic operation, and it is important
to understand if in this case the mapping proposed above is still relevant.
To address this, we consider the standard approach of small-signal analysis 
(SSA),\cite{Chuang:2009} and apply it to both QD and QW lasers.  
We decompose the quantities of
interest,  $X$, into a steady-state $X_0$ and a (small) modulated part 
$\delta X(t)$, 
$X=X_0+\delta X (t)$, and focus on the harmonic modulation  
$\delta X(t)=\mathrm{Re}[\,\delta X(\omega) e^{-i \omega t}]$, 
where $\omega$ is the (angular) modulation frequency. 
The response function, which characterizes the dynamic operation 
including the laser bandwidth, 
an important figure of merit, is given by
\begin{equation}
R(\omega)=\left | \delta S(\omega)/ \delta J(\omega) \right|.
\label{eq:R}
\end{equation}
It is convenient to consider the normalized frequency
response function\cite{Chuang:2009}
\begin{equation}
\left|\frac{R(\omega)}{R(0)}\right|_\mathrm{QW}
=\frac{\omega_R^2}{[(\omega_R^2-\omega^2)^2+\omega^2\gamma^2]^{1/2}}, 
\label{eq:RQW}
\end{equation}
where $\omega_R^2\approx g_0 S_0 / \left[ \tau_\mathrm{ph}(1+\epsilon S_0) \right]$
is the relaxation oscillation frequency, and $\gamma$ is a damping factor.\cite{Fowles:2005,wR} 
The functional form of Eq.~(\ref{eq:RQW}) is  the same as for amplitude
of a harmonically-driven damped harmonic oscillator.\cite{Fowles:2005} 
It is useful to express the damping factor as 
\begin{equation}
\gamma\approx2Bn_T+K\,[\omega_R^2/(2\pi)]^2, 
\label{eq:gamma}
\end{equation}
where the $K$-factor
\begin{eqnarray}
K&\approx&4\pi^{2}\left(\tau_\mathrm{ph}+\epsilon/g_0 \right) \label{eq:K} \label{eq:K}
\end{eqnarray}
is an important characteristic parameter  
that determines the high-speed operation limit of lasers. 
In the above equations, we assume $\epsilon \ll g_0/(2Bn_T)$; 
the exact forms are given in Appendix B.

The bandwidth of the laser, $\omega_\mathrm{3dB}$ (see Appendix B), 
is the frequency at which the square of $|R(\omega)/R(0)|$
in Eq.~(\ref{eq:RQW}) is reduced by 3 dB. 
$\omega_\mathrm{3dB}$ and $\omega_R$ are functions of 
the steady-state injection $J_0$, and they coincide for the maximum bandwidth
$\omega_\mathrm{3dB}^\mathrm{max}$.
Commonly, the peak position $\omega^2_\mathrm{peak}=\omega_R^2-\gamma^2/2$
in the response function is approximated as $\omega_R$ 
(when $\omega_R\gg\gamma$, i.e., weak damping),
while the bandwidth in QW lasers can be related to 
$\omega_\mathrm{peak}$ and $\omega_R$,\cite{Coldren:1995,Asryan2010:APL} 
\begin{equation}
\omega^2_\mathrm{3dB}=\omega^2_\mathrm{peak}+(\omega_\mathrm{peak}^4+\omega_R^4)^{1/2}.
\label{eq:O3QW}
\end{equation}
The maximum bandwidth
is attained for $\omega_R^2=\gamma^2/2$ to 
give a monotonic decrease of response function defined in Eq.~(\ref{eq:RQW}).
\begin{figure}
\includegraphics[scale=0.6]{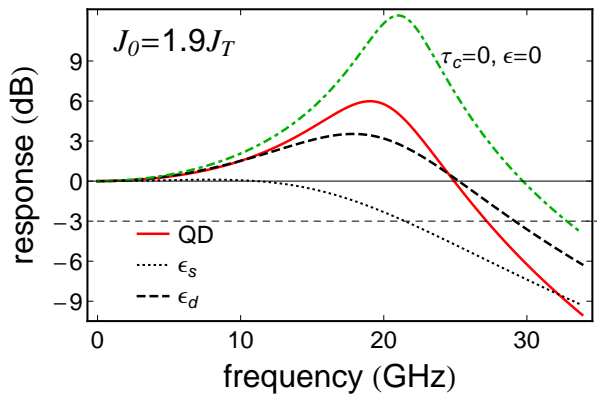}
\caption{(Color online) 
Small-signal analysis for QD and QW lasers, given by the square of the
normalized  frequency response function.  
The solid line represents QD laser response ($\tau_c=2$ ps; see Table \ref{tab:01}), 
while for QW lasers, the dashed and dotted 
lines correspond to static and dynamic gain compression factors 
$\epsilon_{s}=1.62\times10^{-14}$~cm$^3$ and 
$\epsilon_{d}=6.39\times10^{-15}$~cm$^3$, respectively.
The response for $\epsilon=0$ (dot-dashed), 
identical to the response of the QD laser for $\tau_c=0$, 
is shown for comparison.
\label{fig:04}}
\end{figure}
For QD lasers the response function can be related to
its QW counterpart in Eq.~(\ref{eq:RQW}), under the assumption
of $\omega_r'\ll1/\tau_c'$, where $\omega_r'$ ($\gamma_\mathrm{QD}$) for a QD laser
corresponds to $\omega_R$ ($\gamma$) for a QW laser, 
and $\tau_c'\approx\tau_c/(1-f_{q0})$ is the effective capture time.
In this regime, often realized experimentally,  we obtain (more general 
expressions are given in Appendix B) 
\begin{equation}
\left|\frac{R(\omega)}{R(0)}\right|_\mathrm{QD}
\approx(1+\omega^2\tau_c'^2)^{-1/2} \times
\left|\frac{R(\omega)}{R(0)}\right|_\mathrm{QW}.
\label{eq:RQD}
\end{equation}
Analogously to QW lasers, the bandwidth for QD lasers can be obtained from the equation 
\begin{equation}
(1+\omega_\mathrm{3dB}^2\tau_c'^2)[(\omega_r'^2-\omega_\mathrm{3dB}^2)^2
+\omega_\mathrm{3dB}^2\gamma_\mathrm{QD}^2]=2\omega_r'^4, 
\label{eq:O3QD}
\end{equation}
which in the limit of $\tau_c \to 0$, 
recovers the QW behavior, determined by Eq.~(\ref{eq:O3QW}).
Our REs for the QD laser resemble those for separate confinement heterostructure 
(SCH) lasers,\cite{Nagarajan1992:JQE} except that for QD lasers it 
is important to consider the Pauli exclusion principle. The Pauli factor, which appears 
in the form of $1-f_{\eta}$, not only reduces the steady-state photon 
density, but also significantly suppresses the modulation response. 
Similarly to the
corresponding capture in SCH lasers,\cite{Chuang:2009} the contribution 
of $\tau_c$ to the QD modulation response function is responsible for low-frequency roll-off (i.e., negative slope of the response). 
When $\tau_c$ is so large that the roll-off is dominant, 
the maximum bandwidth attained is $\sim1/\tau_c'$. 
Our results imply that to maximize the QD laser dynamic response, the capture time
should be sufficiently short, $\tau_c<10$ ps, consistent with a 
previous study of QDs.\cite{Ohnesorge1996:PRB,Heitz1997:PRB,Sugawara1997:APL,%
Muller2003:APL,Ishida2004:APL} 
As mentioned in Sec.~\ref{SSM}, 
it is required that the mapped QW laser REs recover the dynamics of QD lasers. 
\begin{figure}
\includegraphics[scale=0.6]{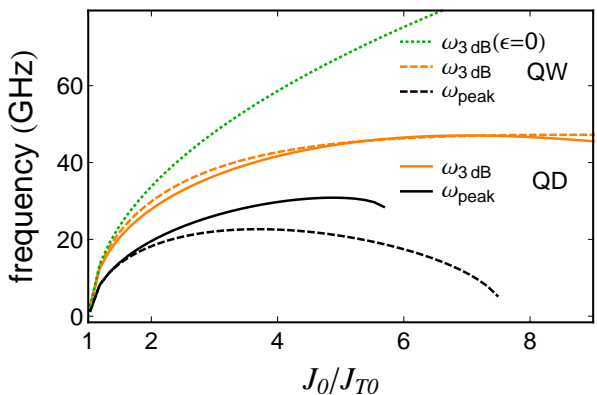}
\caption{(Color online) 
Injection dependence of characteristic frequencies obtained from
small signal analysis:
Gray (orange) and black lines 
show bandwidth ($\omega_\mathrm{3dB}$) and peak frequency ($\omega_\mathrm{peak}$), respectively.
Solid and dashed lines represent QW and QD lasers, respectively. 
The dotted line is the bandwidth of a mapped QW laser for $\epsilon=0$.
\label{fig:05}}
\end{figure}
While our goal is not to fully recover a detailed dynamic response of QD lasers,  
we do require that the maximum bandwidth $\omega_\mathrm{3dB}^\mathrm{max}$, as
the key figure of merit characterizing dynamical operation, coincides for
QD and QW lasers. This can be achieved through a $K$-factor
that defines the maximum frequency for QW lasers and depends on $\epsilon$.
The mapping is then realized by following the
same matching procedure and conditions (i)-(iii) described in Sec.~\ref{SSM}, 
while for the previous condition (iv) is now replaced by the expression for
$\epsilon_d$ which reflects the matching of the maximum bandwidth from 
Eq.~(\ref{eq:K}),  
\begin{equation}
\epsilon_{d}\approx g_0 \left(\sqrt{2}/\omega_\mathrm{3dB}^\mathrm{max}-\tau_\mathrm{ph} \right),
\label{eq:eD}
\end{equation}
where the subscript $d$ refers to the dynamical response with the corresponding
value which does not need to coincide with the value obtained in the steady-state mapping,
i.e., $\epsilon_s$. The maximum bandwidth $\omega_\mathrm{3dB}^\mathrm{max}$ is obtained
from the QD laser REs; Eq.~(\ref{eq:eD}) is valid for $\omega_\mathrm{3dB}^\mathrm{max}\gg2B n_T$ 
and $\epsilon_d\ll g_0/(2Bn_T)$. 
Equation~(\ref{eq:eD}) gives a less than 3\% error with $\tau_c=2$ ps, 
compared to exact calculation [see Eq.~(\ref{eq:Omax}) in Appendix B]; 
however, for mapping over a wide 
range of $\tau_c$, we used the general expressions presented in Appendix B. 

To examine differences between the two mapping procedures, in Fig.~\ref{fig:04}
we compare the response function of a QD laser to response functions calculated for QW
lasers from both steady-state and dynamical-response mapping at a given
injection ($J_0=1.9 J_T$).
In the limit $\tau_c=0$, REs for QD lasers reduce to REs for QW lasers with  
$\epsilon=0$.  We see qualitative similarities for finite
$\epsilon$ and $\tau_c$ which are both detrimental and cause bandwidth suppression. 
In small-signal analysis, the calculated QW laser response function shows 
a wider spread and different slope in the tail
than the one for QD lasers. As can be seen from Fig.~\ref{fig:04}, 
use of the gain compression factor obtained from the steady-state
mapping, $\epsilon_s$, provides a poor approximation to the response function
for  a QD laser. The agreement is considerably better 
when a much smaller gain compression factor from dynamical-operation mapping, 
$\epsilon_d$, is used instead. 
We see that the QD
and QW response functions are nearly indistinguishable
up to $\sim$ 10 GHz (the maximum bandwidths are matched for
higher currents).

To assess the quality of the dynamic-operation mapping, in Fig.~\ref{fig:05}
we show the injection dependence of the bandwidth $\omega_\mathrm{3dB}$ 
and the peak position $\omega_\mathrm{peak}$ 
for the QD and QW lasers.
It is remarkable that 
the mapping, only intended to match the maximum bandwidth
between the QD and QW lasers, 
yields  a very good agreement for the bandwidth dependence on injection.  
Both QD and QW cases reveal a nonmonotonic behavior up to 
$J\sim6 J_{T0}$.  While Fig.~\ref{fig:04} shows a very similar 
peak position for QD and QW lasers, from Fig.~\ref{fig:05} we can infer that 
this occurs typically only close to the threshold injection.
The discontinuity of $\omega_\mathrm{peak}$ for the QD laser (black solid curve) is due to 
low-frequency roll-off. 
As a result of the interplay of $\tau_c'$ and $\omega_r'$, shown in Eq.~(\ref{eq:O3QD}),
$\omega_\mathrm{peak}=0$ 
(analogous to an overdamped harmonic oscillator\cite{Fowles:2005}) 
only above injection $J_0\sim5.7J_T$.

\section{Steady-State vs Dynamic-Operation Gain Compression}
\label{SSM vs DOM}

\begin{figure}
\includegraphics[scale =0.6] {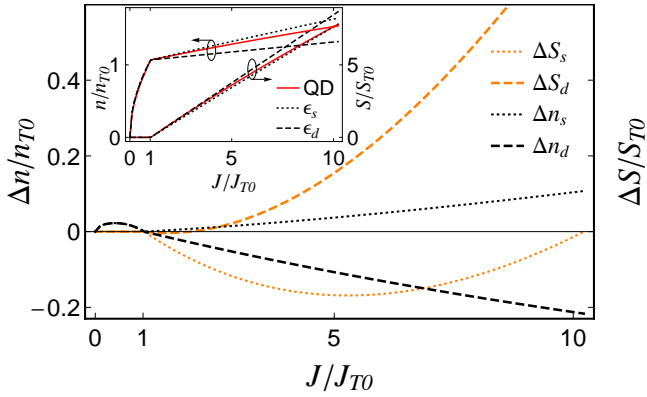}
\caption{(Color online) 
Deviations of QD laser photon ($\Delta S$) and carrier ($\Delta n$) densities
from those obtained by mapped QW lasers with $\epsilon_{s}$ (dotted) 
and $\epsilon_{d}$ (dashed).  Black [gray (orange)] lines are deviations in carrier 
(photon) densities (same scale as for $\Delta n/n_{T0}$). 
Note that $10J_T \approx 10.23J_{T0}$. 
Inset: Carrier and photon densities for QD and QW lasers with $\epsilon_s$ 
and $\epsilon_d$ are shown by solid, dotted, and dashed 
lines, respectively. Note the different vertical scales.
\label{fig:06}}
\end{figure}

In the preceding two sections we have formulated steady-state and dynamic-operation 
mapping  and showed that, with the corresponding change in $\epsilon$, there are 
considerable differences when it comes to small-signal analysis. We now
examine if these differences, between choice of $\epsilon_{s}$ and $\epsilon_{d}$,
also persist in the steady-state regime. In Fig.~\ref{fig:06} 
we consider light-injection and 
carrier density-injection characteristics. 
The light intensity at $J=10J_T$ for QW laser dynamic mapping ($\epsilon_d$) is about 
10 $\%$ higher than for QD and QW lasers steady-state mapping ($\epsilon_s$).
The light intensity at $J=10J_T$ is set to be the same 
for QD and QW laser
with $\epsilon_s$ chosen according to the matching condition (iv) in Sec.~\ref{SSM}. 
The carrier density of the QW lasers is
noticeably different from the of the QD laser. Typically the relative differences in carrier density 
are more pronounced than in the light intensity (see Fig.~\ref{fig:06} inset).
Since, generally,  $\epsilon_s > \epsilon_d$,  
a higher light intensity is maintained for $\epsilon_d$ at the same injection 
by consuming more carriers in the active region through stimulated recombination. 
Therefore, at $J=10J_T$, the carrier density of QW lasers with $\epsilon_d$ (gray solid) 
is about 30 $\%$ lower than that of QW lasers with $\epsilon_s$ (black solid).

\begin{figure}
\includegraphics[scale =0.6] {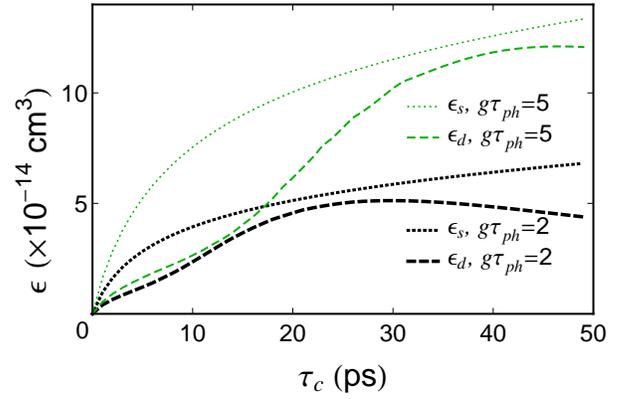}
\caption{(Color online) 
Gain compression factors  $\epsilon_s$ (dotted) and $\epsilon_d$
(dashed), obtained from two different mapping procedures. 
Thick (black) and thin [gray(green)] lines represent $g \tau_\mathrm{ph}=2$ and 5, respectively,
for $\tau_\mathrm{ph}=2$ ps.
\label{fig:07}}
\end{figure}

Recognition of the correspondence between the increasing capture time
in QD lasers and the increasing gain compression factor in QW lasers
was the basis for both the steady-state and dynamic mapping. 
In the previous plots (see Figs.~\ref{fig:03} and \ref{fig:04})
we focused on a modest capture time ($\tau_c = 2$ ps). 
In Fig.~\ref{fig:07}, 
$\epsilon_{s}$ and $\epsilon_{d}$ of 
mapped QW lasers 
are plotted as functions of capture time of QD lasers, for different 
gain coefficients.
These results show that, when the two mappings are compared, 
$\epsilon_s$ is always greater than $\epsilon_d$, which leads to an 
excessive  suppression of dynamic response when the steady-state 
mapping of QD lasers is implemented. 
While $\epsilon_s$ shows a monotonic increase with $\tau_c$,
there is a nonmonotonic variation of $\epsilon_d$. 
In particular, $\epsilon_d$ has a local maximum at $\tau_c\sim30$ (43) ps 
for $g\tau_\mathrm{ph}=2$ (5) and starts to 
decline for large $\tau_c$. 
This unexpected behavior of $\epsilon_d$ reflects a rapid decrease of the 
mapped QW laser gain $g_0$. 
The maximum bandwidth $\omega_\mathrm{3dB}^\mathrm{max}$ of a QD laser 
decreases with 
increasing $\tau_c$, and in maintaining the same value of
$\omega_\mathrm{3dB}^\mathrm{max}$
within QW laser REs, $\epsilon_d$ and $g_0$ play an important role. 
As $\tau_c$ grows, $g_0$ and $B$ respectively decline and increase 
according to the conditions (ii) and (iii) 
in Sec.~\ref{SSM}. Beyond $\tau_c \sim 30$ (43) ps,  
$\omega_\mathrm{3dB}^\mathrm{max}$ 
tends to saturate to $2Bn_T$, while the decrease of gain retains its rate. 
As a result, $\epsilon_d$ has to stop rising and even starts to decrease with $\tau_c$ 
to compensate for the rapidly diminishing gain, leading to the
maximum of $\epsilon_d$ in Fig.~\ref{fig:07}. 

\section{Mapping of Spin Lasers}
\label{MOS}

Our preceding analysis of mapping was limited to the absence of injected spin
polarization ($P_J=0$). The more general case of spin lasers ($P_J \neq 0$)
adds complexity to REs, requiring four equations for QW and ten for QD lasers
(see Appendix A). 
For QDs, the added complexity prevents analytical solutions
even in the steady state, making any attempt at directly implementing the 
mapping for spin lasers more challenging. On the other hand, this same 
complexity implies that the prospect of studying QD spin lasers by considering a simpler description for QW spin lasers will be more valuable 
than in the conventional lasers. Moreover, important recent experiments
on QD-based spin lasers\cite{Basu2008:APL,Basu2009:PRL,Saha2010:PRB}  
are described within the formalism of QW spin laser REs and it is not {\em a priori} 
clear how accurate is such a procedure. Typically, these spin lasers are realized 
in a Faraday geometry\cite{Zutic2004:RMP} as vertical-cavity surface-emitting 
lasers (VCSELs).\cite{Yu:2003} The main difference from
commercially available VCSELs is the presence of spin-polarized carriers,
provided by pumping with circularly polarized light or using magnetic contacts
for electrical spin injection.\cite{Rudolph2003:APL, Rudolph2005:APL,%
Holub2007:PRL,Gerhardt2011:APL,Hovel2008:APL,Jahme2010:APL,%
Hovel2007:PSS,Ando1998:APL,%
Hallstein1997:PRB, Soldat2011:APL, Fraser2010:APL}

\begin{figure}[tbh]
\includegraphics[scale =0.6] {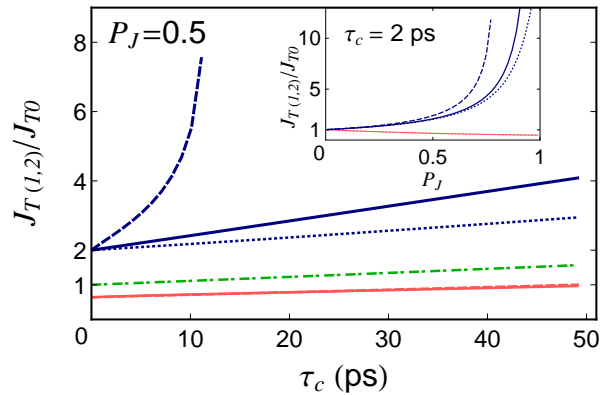}
\caption{(Color online) 
First [gray(red)] and second [black(blue)] thresholds as functions 
of capture time $\tau_c$ for $P_J=0.5$. Solid lines represent the QD spin laser, 
while dotted and dashed lines correspond to QW spin lasers 
with self- and even- compression of gain, respectively. 
For comparison, the threshold of a conventional QD laser ($P_J=0$) 
is shown by the dot-dashed line. Inset: The thresholds as functions of 
injection polarization $P_J$ for $\tau_c=2$ ps. 
\label{fig:09}}
\end{figure}

In spin lasers we consider 
spin-resolved quantities to model different spin projections or helicities 
of light. The total electron or hole density can be written as the sum of the
spin-up (+) and the spin-down ($-$) electron or hole densities,
$n=n_+ + n_-$ and $p=p_+ + p_-$. Analogously, we write the total photon
density as the sum of the positive (+) and negative ($-$) helicities,
${S{\,}={\,}S^+{\,}+{\,}S^-}$.
A generalization
of the optical gain term in Eq.~(\ref{eq:gain}) for QW spin lasers
can be expressed as
\begin{equation}
g_\pm(n_\pm,S^\pm)=g_0(n_\pm+p_\pm-n_\mathrm{tran})/(1+\epsilon^\pm_+ S^+ + \epsilon^\pm_- S^-),
\label{eq:spingain}
\end{equation}
where $g_{\pm}$ is the spin dependent gain which couples to the corresponding 
spin of carriers $n_{\pm}$. The superscript of $\epsilon$ represents the spin of  
coupled carriers, while the subscript represents the corresponding helicity of photons.
Due to the symmetry, 
$\epsilon_{+}^{-}=\epsilon_{-}^{+}=\epsilon_{\mathrm{cross}}$ and 
$\epsilon_{+}^{+}=\epsilon_{-}^{-}=\epsilon_{\mathrm{self}}$. 
The index cross (self)
implies a cross- (self-) compression mechanism of gain. Later in this section (Fig.~\ref{fig:09}), 
we compare the self-compression limit 
($\epsilon_{\mathrm{self}}=2\epsilon$, $\epsilon_{\mathrm{cross}}=0$ ) to 
the even-compression limit ($\epsilon_{\mathrm{self}}= 
\epsilon_{\mathrm{cross}}=\epsilon$).
Each case recovers the spin-unpolarized laser REs for $P_J=0$.

\begin{figure}
\includegraphics[scale =0.6] {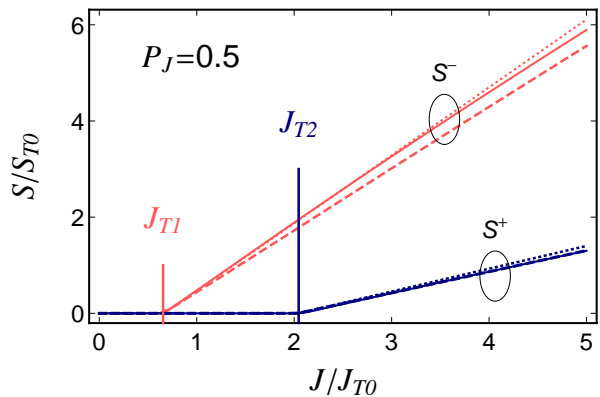}
\caption{(Color online) 
Photon densities of the spin laser are shown for injection $P_J=0.5$.
Solid, dashed, and dotted lines represent QD and QW lasers with $\epsilon_d$ and $\epsilon_s$ 
(given in Table~\ref{tab:01}), 
while gray and black lines represent left ($S^-$) and 
right ($S^+$) circular polarization, respectively. $S$ and $J$ are 
normalized to the values $S_{T0}$ and $J_{T0}$ for $P_J=0$ and 
$\tau_c=0$ ($\epsilon=0$). Vertical lines indicate thresholds for 
majority ($J_{T1}$) and minority ($J_{T2}$) spin carriers.
\label{fig:08}}
\end{figure}

To establish a connection between
QD and QW  spin lasers, we reconsider our mapping
procedure discussed above for $P_J=0$. We focus on the regime of a strong
electron-hole spin asymmetry, shown to lead to maximum threshold 
reduction\cite{Gothgen2008:APL, Vurgaftman2008:APL} 
and desirable dynamical properties of spin lasers,\cite{Lee2010:APL} in which the spin 
relaxation time of holes is much shorter than for the electrons. For example,
in bulk GaAs at room temperature the measured spin relaxation time of holes 
is $\sim 100$ fs,\cite{Zutic2004:RMP} and of electrons it is
$\sim 0.1-1$ ns.\cite{Fabian2007:APS} In spin lasers
it is therefore customary to consider that holes are spin unpolarized. 
Here, for simplicity, we also focus mostly on the infinitely long spin relation times 
for electrons (in the QW, WL,  and QD regions). This limiting case can 
accurately describe recent experiments,\cite{Fujino2009:APL,Ikeda2009:PTL}
in which the spin relaxation time for electrons is not only much longer than 
for holes, but also much longer than the other characteristic timescales 
for the carriers. 

The light-injection characteristics obtained for mapping of QD to QW 
spin lasers with self-compression are shown in Fig.~\ref{fig:08}. 
Several key features of spin lasers that can already be inferred from 
the bucket model in Fig.~\ref{fig:02b} 
are clearly present. With $P_J \neq 0$ the 
thresholds for majority and minority spin ($J_{T1}$ and $J_{T2}$) are
different. Since  $J_{T1} < J_T < J_{T2}$ there is a threshold 
reduction $r$ [recall Eq.~(\ref{eq:r})], as compared to conventional 
lasers. Furthermore, for injection $J_{T1} < J < J_{T2}$
there will be a spin-filtering effect [Eq.~(\ref{eq:d})]; even a 
modest injection leads to fully polarized 
emitted light.\cite{Lee2010:APL,Saha2010:PRB} 
Even though our results have been based on parameters 
identical to the ones used for conventional lasers
(supplemented by the  vanishing hole and infinite electron spin 
relaxation times), we retain a good agreement between QD
and QW lasers, especially near the two thresholds. For example,
in Fig.~\ref{fig:08} within dynamical mapping determined by $\epsilon_d$,
the emitted right circular polarization, $S^+$ [black(blue) dashed line]
is almost indistinguishable for  QD and  QW lasers.  

We further explore the mapping of spin lasers in Fig.~\ref{fig:09}; the inset
shows the evolution of majority and minority thresholds with injection 
polarization, for both even- and self-compression of gain.
There is an excellent agreement of $J_{T1}$ for all $P_J$'s (the three curves overlap) and 
a good agreement of $J_{T2}$ up to $P_J \sim 0.6$, which implies that, 
within practical injection polarization of spin lasers realized
at room temperatures, the proposed mapping works well.  
From the dependence of $J_{T2}$ on $P_J$ in QD lasers
we see that their
behavior falls between those of the QW approximations using self-compression only 
and even-compression. 
The same trend, i.e.,~$J_{T2}$ of a QD being bounded by
the two limiting cases for the gain compression of QW lasers, is 
also shown in the main panel as a function of $\tau_c$ at fixed $P_J$.
The threshold $J_{T2}$ in QW lasers disappears for 
the even-compression approximation, as  can be seen both in the inset 
($P_J \approx 0.83$) and in the main panel 
($\tau_c \approx 10$ ps). 
In contrast, there is no
disappearance of $J_{T2}$ for QW lasers with self-compression.
The high accuracy of $J_{T1}$ mapping is not limited to the specific 
approximation of gain compression
(it is independent of $\epsilon$) 
and persists for a wide parameter range
(in both $P_J$ and $\tau_c$). As a consequence, the threshold reduction
[Eq.~(\ref{eq:r})] of QD lasers is well approximated by mapping to QW
lasers. The spin-filtering regime, given by Eq.~(\ref{eq:d}), is
present in both QD and QW lasers, but its dependence on $J_{T2}$ implies
less accuracy at higher values of $P_J$ and $\tau_c$, while the latter range is 
experimentally less relevant. 
We note that in Fig.~\ref{fig:09}, only $\epsilon_s$ is used for the calculation since
the results are insensitive to the difference between $\epsilon_s$ and $\epsilon_d$.
As mentioned above, $J_{T1}$ is independent of $\epsilon$, and 
the difference in $J_{T2}$ due to the discrepancy between $\epsilon_s$ and $\epsilon_d$
is less than 2\%.

\section{Small-Signal Analysis for Spin-Lasers}
\label{SSA}

Motivated by the early steady-state experiments on spin lasers, it was
predicted that the observed threshold reduction could also lead to
desirable dynamic operation and the bandwidth 
enhancement.\cite{Lee2010:APL,Banerjee2011:JAP} Recent advances in electrical and
optical spin injection\cite{Saha2010:PRB,Fujino2009:APL, Ikeda2009:PTL,Iba2010:APL,Bull2004:PSPIE,Hanbicki2003:APL,Salis2005:APL,%
Zega2005:PRL,Crooker2009:PRB,Li2009:APL} 
suggest versatile opportunities for the  
modulation of spin lasers. In previous work on QW spin lasers we considered 
amplitude and polarization modulation (AM, PM).\cite{Lee2010:APL}

AM for a steady-state polarization implies $J_+ \neq J_-$ (unless 
$P_J=0$ when AM recovers its standard form for conventional lasers),
\begin{equation}
\mathrm{AM}: \:
J=J_0+\mathrm{Re}[\,\delta J(\omega) e^{-i\omega t}], \quad  P_J = P_{J0}. 
\label{eq:AM}
\end{equation}
As in the steady-state analysis, 
$P_J \neq 0$ leads to unequal threshold currents $J_{T1}$ and $J_{T2}$, 
apparent already from the bucket model in 
Fig.~\ref{fig:02b}.
Such a modulation can be contrasted with PM,
which also has $J_+ \neq J_-$, but
$J$ remains constant:\cite{const}
\begin{equation}
\mathrm{PM}: \:
J=J_0, \quad P_J=P_{J0}+\mathrm{Re}[\,\delta P_J(\omega) e^{-i\omega t}].
\label{eq:PM}
\end{equation}
It was recently shown that a similar PM scheme could enable high-performance 
spin-communication schemes with an effective information transfer rate that exceeds
currently available realizations by several orders of magnitude.\cite{Dery2011:APL}

\begin{figure}[tbh]
\includegraphics[scale =0.85] {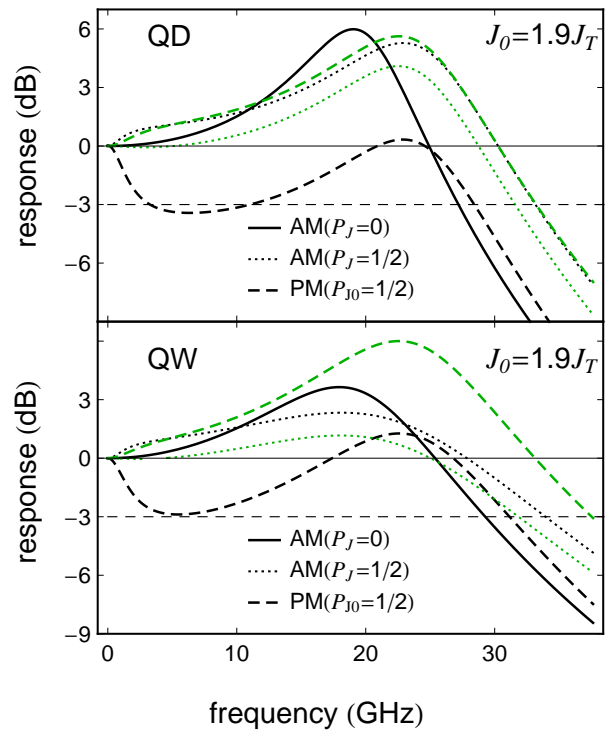}
\caption{(Color online) 
The square of the normalized  frequency response function of   
QD (upper) and QW (lower) spin lasers.
Solid, dashed, and dotted lines represent amplitude modulation (AM) for $P_{J0}=0$ and 
$0.5$ and polarization modulation (PM) for $P_{J0}=0.5$, respectively.
Injection $J_0$ is fixed at $1.9 J_T$. Gray (green) lines represent 
finite electron spin relaxation time $tau_s=200$ ps 
for AM (dotted) and PM (dashed), respectively.
\label{fig:10}}
\end{figure}

We generalize the small-signal analysis outlined in
Sec.~IV and compare our results for conventional lasers with
those for spin lasers, using $\epsilon_{d}$ and 
self-compression. The response function can be generalized as 
$R_{\pm}(\omega)=|\delta S^{\mp}(\omega)/\delta J_{\pm}(\omega)|$ for spin lasers 
and it reduces to $R(\omega)$ for $P_{J0}=0$ (AM).
In Fig.~\ref{fig:10}, we consider both AM and PM,
choosing $P_J=0.5$ and the injection $J_{T1} <J_0=1.9 J_T <J_{T2}$, which lies in
the spin-filtering regime. As in the previous studies of QW spin lasers, 
we see that both AM and PM can lead to enhanced bandwidth 
($\omega_\mathrm{3dB}$), as compared to conventional lasers (birefringence in
spin lasers could provide additional paths to enhanced 
bandwidths\cite{Gerhardt2011:APL, Jahme2010:APL}). 
The shape of the frequency response of spin lasers in Fig.~\ref{fig:10}
is significantly modified from what was previously obtained in 
Ref.~\onlinecite{Lee2010:APL} due to the large $\epsilon_d$. 
This is particularly pronounced for PM, which shows 
a low-frequency roll-off.  Despite the fact that the maximum  $\omega_\mathrm{3dB}$ for 
a spin laser is enhanced, the useful frequency range for PM may be reduced
due to the low-frequency roll-off before the response peak. 
From Fig.~\ref{fig:10} we see that the dynamic mapping of spin lasers 
preserves qualitative features of the 
frequency response, and thus insight into the QD spin lasers can be sought
from the much simpler QW REs. 

\section{conclusions}
\label{CON}

We have formulated a systematic approach which allows mapping of 
QD to QW lasers and thus reduces the complexity of 
the QD laser description based on rate equations. The key observation
to establish this mapping is that the influence of finite $\tau_c$ on the
operation of  QD lasers can be approximated well by a suitable choice
of the gain compression factor $\epsilon$ in the simpler QW lasers. 

\begin{figure}[tbh]
\includegraphics[scale =0.45] {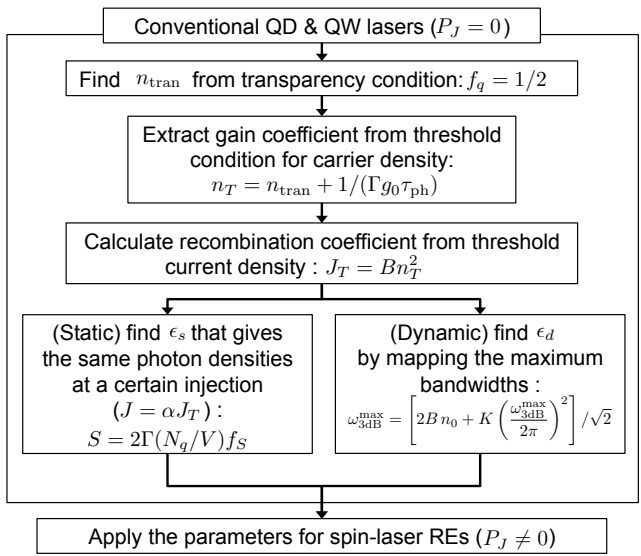}
\caption{
Schematic representation of the mapping.
\label{fig:11}}
\end{figure}
However, the choice of how $\tau_c$ should be related to $\epsilon$ is not unique;
we find noticeable differences between the mappings 
of either steady-state or dynamic operation of lasers, corresponding to the
respective values of the gain compression factors $\epsilon_s$ and $\epsilon_d$. 
The mapping procedure, schematically outlined in the Fig.~\ref{fig:11}, can be realized 
either analytically or numerically. 
The steady-state mapping preserves well the behavior of a QD laser near its threshold,
for both conventional and spin lasers. In the latter case, for an arbitrary injection 
spin polarization, the majority threshold is particularly accurate, further justifying 
the use of QW spin lasers REs for threshold reduction.\cite{Rudolph2003:APL,%
Holub2007:PRL,Gothgen2008:APL,Oszwaldowski2010:PRB,Vurgaftman2008:APL}. 
When the dynamic-range mapping is considered, we focus on preserving the 
\emph{maximum} bandwidth of QD and QW lasers, since their detailed behavior 
can display considerable 
differences, including low-frequency roll-off in the modulation response.\cite{Fathpour2005:JPD}
Additional motivation for this approach is that the bandwidth itself depends on the injection
level and therefore it would not be as useful as a quantity to be matched in the mapping.

The growing interest in QD lasers and the increasing number of materials used
for the active region [such as colloidal QDs (Refs. \onlinecite{Klimov2007:N,Scholes2008:AFM,%
Beaulac2008:AFM,Stern2005:PRB})] 
provides  a further motivation to construct a mapping discussed in this work. Since
the mapping is not limited to conventional lasers, it can also be used to
guide further developments of QD spin lasers. The presence of QDs in the 
active region leads to reduced influence of the spin-orbit coupling,\cite{Fabian2007:APS}
resulting in a longer spin relaxation time which improves lasing properties, giving a 
lower threshold and enhanced bandwidth. 
Detailed knowledge of the structures used in recent experiments on QD spin lasers,%
\cite{Basu2008:APL,Basu2009:PRL,Saha2010:PRB} 
would allow us to apply the mapping outlined above and examine how it is related 
to a description based on densities rather than occupancies.\cite{Saha2010:PRB} 

Several assumptions of the present mapping could be relaxed. 
To allow a more general RE description of QD lasers, it might be possible to 
include explicitly a finite gain compression factor into QD laser REs.%
\cite{Qasaimeh2009:JLT}
The expected change in the mapping procedure would be an appropriately
rescaled (enhanced) $\epsilon$ for QW lasers, playing the combined 
role of the $\epsilon$ of the QD lasers and the finite $\tau_c$.  
With further studies of self- and even-compression mechanisms, it would be
possible to more accurately model the gain compression with 
appropriately weighted contributions of the two mechanisms (reflected
in the matrix structure of $\epsilon_\pm^\pm$). 
Future generalizations of the mapping procedure could also consider finite spin relaxation
times of holes. While the spins of holes in bulk GaAs  at 300 K can very accurately be 
treated as being lost instantaneously (approximately 3-4 orders of magnitude faster than
the spin of electrons),\cite{Zutic2004:RMP} in QDs the asymmetry of 
spin relaxation times for electrons and holes should be reduced.\cite{Fabian2007:APS}   

In a future work it would also be interesting to explore other forms of mapping
procedures that could establish similarities between spin lasers and phase
transitions in magnetic systems. Such a consideration would generalize
what is already known for conventional lasers, linked to Ising 
ferromagnets,\cite{Degiorgio1970:PRA,Degiorgio1976:PT}
and explain how the spin imbalance inherent to spin lasers can, 
through suitable mapping, be related to a more complex magnetic behavior.  

\section{Acknowledgments}

This work was supported by the NSF-ECCS 1102092, NSF-ECCS CAREER, U.S. ONR,
AFOSR-DCT, DOE-BES, NSF-NRI NEB 2020, and SRC. 
We thank H. Dery for valuable discussions. 

\appendix
\section{}

For QW spin lasers, the REs given by Eqs.~(\ref{eq:REn}) and (\ref{eq:RES}) are generalized as
\begin{eqnarray}
dn_{\pm}/dt&=&J_{\pm}-g_{\pm}(n_{\pm},S^{\mp})S^{\mp}
-R^{\pm}_\mathrm{sp}\mp F,\label{QWspin}\\
dS^{\mp}/dt&=&\Gamma g_{\pm}(n_{\pm},S^{\mp})S^{\mp}
+\Gamma \beta R^{\pm}_\mathrm{sp}-S^{\mp}/\tau_\mathrm{ph},\nonumber
\end{eqnarray}
where the $+/-$ subscript (superscript) represents the corresponding electron spin 
(photon helicity). In Eq.~(\ref{QWspin}) an additional term, vanishing for 
$P_J=0$ in  conventional lasers, corresponds  to spin 
relaxation $F=(n_{\pm}-n_{\mp})/\tau_s$, 
where $\tau_s$ represents the electron spin relaxation time $\tau_{sn}$. 
Spontaneous recombination is written as $R^\pm_\mathrm{sp}=2Bn_\pm p_\pm$.  
The instantaneous hole spin relaxation $\tau_{sp}\to0$ allows us to write the hole 
density in terms of electron densities as $p_+=p_-=p/2=(n_++n_-)/2$, 
which results in $R^{\pm}_\mathrm{sp}=Bn_{\pm}(n_{+}+n_{-})$, with the 
assumption of charge neutrality.

\begin{figure}[tbh]
\includegraphics[scale=0.5]{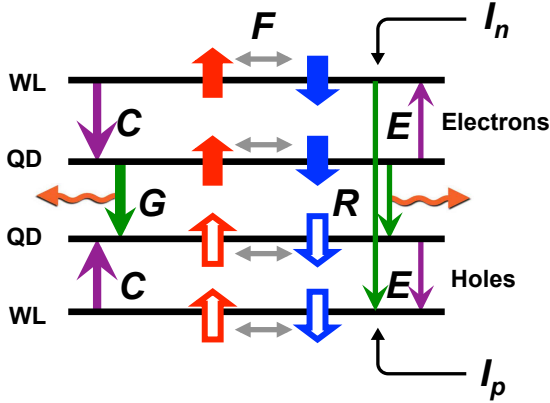}
\caption{
(Color online) Processes in QD spin lasers described by Eqs.~(\ref{QDspin}).
QD (WL) represents the level for the quantum dot (wetting layer).
The upper two levels are for electrons and the lower levels represent levels for holes.
The thick vertical arrows show the carrier spin filled for electrons, empty for holes. 
The thin arrows depict carrier injection $I$, capture $C$, escape $E
$, spin relaxation $F$, and stimulated $G$ and spontaneous $R$ recombination 
in QDs and QWs (thickness indicates relative rates). 
The subscripts $n$ and $p$ represent the electron and 
hole contributions, respectively. Wavy arrows depict photon emission.
\label{fig:12}}
\end{figure}

An  important difference between QW and QD spin laser REs is that 
only QD spin laser REs have explicit terms for hole occupancies. 
In QW spin lasers, 
hole densities can be easily replaced by electron densities as discussed above 
($\tau_{sp}\to0$). 
However, for QDs, unlike QWs, the ultrafast spin relaxation time for holes 
($\tau_{spw},\tau_{spq}\to0$), does not lift the explicit hole
density dependence of QD REs. This makes it more difficult to analytically
study QD spin lasers even in the steady state. A generalization of the QD 
Eqs.~(\ref{eq:fw})-(\ref{eq:fS})  for spin lasers is 
\begin{eqnarray}
df_{w \alpha \pm}/dt&=&I_{\alpha \pm}-C_{\alpha \pm}
+\frac{2}{\kappa_\alpha}E_{\alpha \pm}-R_{w \pm}\mp F_{w\alpha}, \nonumber\\
df_{q \alpha \pm}/dt&=&\frac{\kappa_\alpha}{2}C_{\alpha \pm}-E_{\alpha \pm}
-R_{q \pm}-G_{\pm}\mp F_{q\alpha},\nonumber\\
df_{S\mp}/dt&=&G_{\pm}+\beta R_{q\pm}-f_{S\mp}/\tau_\mathrm{ph},
\label{QDspin}
\end{eqnarray}
where $\alpha=n,\,p$ denotes electrons and holes, respectively.
$I$, $C$, $E$, $G$, and $R$ represent injection, capture, escape, 
the stimulated and spontaneous emission, respectively, as in unpolarized REs, while
$F$ represents spin relaxation. The level scheme of a QD spin laser is shown 
in Fig.~\ref{fig:12}. Since the occupancies satisfy $0\le f_{(w,q)} \le1$, 
the spin-polarized 
occupancies are defined as 
$f_{w\alpha\pm}=\bar{n}_{w\alpha\pm}/(N_{w\alpha}/2)$,
$f_{q\alpha\pm}=\bar{n}_{w\alpha\pm}/N_{q}$ and
$f_{S\pm}=S_{\pm}/N_{q}$, which are different by a factor of 2 from 
Eqs.~(\ref{eq:fw})-(\ref{eq:fS}). 
The carrier injection, capture, and escape are 
$I_{\alpha\pm}=j_{\alpha\pm}(1-f_{w\alpha\pm})$, 
$C_{\alpha\pm}=f_{w\alpha\pm}(1-f_{q\alpha\pm})/\tau_{c\alpha}$, 
and $E_{\alpha\pm}=f_{q\alpha\pm}(1-f_{w\alpha\pm})/\tau_{e\alpha}$, 
where the injection $j_{\alpha\pm}=(1\pm p_{j\alpha})j_\alpha$ 
can be expressed via the corresponding spin polarization 
$p_{j\alpha}=(j_{\alpha+}-j_{\alpha-})/(j_{\alpha+}+j_{\alpha-})$.
The stimulated and spontaneous emission are
$G_\pm=g(f_{qn\pm}+f_{qp\pm}-1)f_{S\pm}$ and 
$R_{\eta\pm}=b_{\eta}f_{\eta n\pm}f_{\eta p\pm}$, respectively, 
where $\eta =w, q$, and $b_{\eta}$ is the recombination rate.
The spin relaxation term is
$F_{\eta\alpha}=(f_{\eta\alpha+}-f_{\eta\alpha-})/\tau_{s\alpha\eta}$,
where $\tau_{s\alpha\eta}$ is the spin relaxation time.
In this paper, we assume $\tau_{c\alpha}=\tau_c$, $\tau_{e\alpha}=\tau_e$, 
$\tau_{sp\eta}=0$, $\tau_{sn\eta}=\tau_s$, $\beta=0$, $j_\alpha=j$ 
and $p_{j\alpha}=P_J$.

\section{}

A linearization of the QW laser REs Eqs.~(\ref{eq:REn}) and (\ref{eq:RES}),
under a small modulation, leads to the equations for small-signal 
analysis,\cite{Chuang:2009}
\begin{eqnarray}
\begin{bmatrix} A_1-i \omega & A_2 \\ -A_3 & A_4-i \omega \end{bmatrix}
\left[ \begin{array}{c} \delta n \\ \delta S \end{array} \right] 
= \left[ \begin{array}{c} \delta J \\  0 \end{array} \right],
\end{eqnarray}
where the positive matrix elements $A_1$, $A_2$, $A_3$, and $A_4$ are defined as
\begin{eqnarray}
&A_1=2 B n_0+\frac{g_0}{1+\epsilon S_0}S_0,&A_2= \frac{g_0(n_0-n_{\mathrm{tran}})}
{(1+\epsilon S_0)^2}, \label{eq:As}\\
&A_3=\frac{\Gamma g_0(n_0-n_{\mathrm{tran}})S_0}{(1+\epsilon S_0)},&A_4=\frac{1}
{\tau_\mathrm{ph}}-\frac{\Gamma g_0(n_0-n_{\mathrm{tran}})}{(1+\epsilon S_0)^2},
 \nonumber
 \end{eqnarray}
expressed in terms of the quantities introduced in discussion of  
Eqs.~(\ref{eq:REn}) and (\ref{eq:RES}),
as well as their steady-state solutions
$n_0$ and $S_0$ at 
$J=J_0$ . We can obtain the normalized frequency response function as defined in 
Eq.~(\ref{eq:RQW}) with relaxation oscillation frequency $\omega_{R}$ and 
damping factor $\gamma$, given by
\begin{eqnarray}
&\omega_R^2=\frac{g_0 S_0}{\tau_\mathrm{ph}(1+\epsilon\,S_0)}(1+2Bn_0 \epsilon
/g_0)\label{eq:wr2}
\end{eqnarray} and 
\begin{eqnarray}
&\gamma=2Bn_0 + K\,[\omega_R^2/(2\pi)]^2,\label{eq:gamma}
\end{eqnarray}
where $K=(\tau_\mathrm{ph}+\epsilon/g_0)/(1+2Bn_0\epsilon/g_0)$  
is the so-called more precise definition of $K$ factor without approximations. A widely used approximation above 
the threshold, $2B n_0\approx 2B n_T$,\cite{Chuang:2009} 
can also be accurately applied for our mapping. 
On the other hand,  the term $2Bn_0\epsilon/g_0$ 
in Eq.~(\ref{eq:wr2}) is often ignored,\cite{Chuang:2009} but has to be retained
for our purposes of implementing a mapping [Eqs.~(\ref{eq:As})-(\ref{eq:gamma})] with $\epsilon_{s, d},$ 
several orders of magnitude greater than the typical compression factors in QW lasers. 
We therefore use an exact expression for $\omega_R$ and $K$ with finite $\epsilon$
(while considering $\beta=0$ limit).
The bandwidth $\omega_\mathrm{3dB}$ (a function of  injection through $\omega_R$) 
is defined as a frequency that reduces the normalized response function to $1/\sqrt{2}$, 
determined by the equation 
\begin{eqnarray}
(\omega_{R}^{2}-\omega_\mathrm{3dB}^2)^2+\omega_\mathrm{3dB}^2\gamma^2=2\,\omega_{R}^4,
\label{eq:3dB}
\end{eqnarray}
which yields Eq.~(\ref{eq:O3QW}) as its the solution. $\omega_\mathrm{3dB}$ 
is a maximum when the denominator of the normalized response function 
[Eq.~(\ref{eq:RQW})] 
monotonically increases under the condition 
\begin{eqnarray}
\omega_R^2-\gamma^2/2 = 0.\label{eq:C3dB}
\end{eqnarray}
The bandwidth coincides with $\omega_R$ [Eq.~(\ref{eq:O3QW})]
when the condition of Eq.~(\ref{eq:C3dB}) is satisfied, which  
can be written as
\begin{eqnarray}
\omega_\mathrm{3dB}^\mathrm{max}=\left[2B\,n_0+K\left(\frac{\omega_\mathrm{3dB}^\mathrm{max}} 
{2 \pi}\right)^2\right]/\sqrt{2},\label{eq:Omax}
\end{eqnarray}
and we can find the $K$ factor as a function of the maximum bandwidth.
For the dynamic mapping, we substitute for $\omega_\mathrm{3dB}^\mathrm{max}$ the maximum 
bandwidth obtained from the QD laser REs. Once the $K$-factor is found, 
its definition leads us to $\epsilon_d$. 
With the approximations $\omega_\mathrm{3dB}^\mathrm{max}\gg2Bn_0$ and 
$2Bn_0\epsilon/g_0\ll 1$, Eq.~(\ref{eq:Omax}) recovers Eq.~(\ref{eq:eD}). 

One can implement a similar SSA for QD laser REs.  
However, the Pauli blocking terms with the existence of the WL increase 
the complexity 
so that the corresponding response function has a less transparent form.
 For $\beta=0$, the SSA  equations are 
\begin{eqnarray}
\begin{bmatrix} a_1-i\omega & -a_2 & 0 \\
-a_3 & a_4-i\omega & a_5 
\\ 0 & 0 & -a_6-i\omega
\end{bmatrix} \left[ 
\begin{array}{c}
\delta f_w \\ \delta f_q \\ \delta f_S
\end{array} \right] 
=\left[ 
\begin{array}{c} a_7 \delta j \\ 0 \\0 
\end{array} \right], \label{eq:QDSSA}
\end{eqnarray}
where $a_i,$ $i=1,\ldots,7$,  
are positive and defined as
\begin{eqnarray}
a_1&=&j_0+\frac{1-f_{q0}}{\tau_c}+
\frac{2}{\kappa}\frac{f_{q0}}{\tau_e}+2 b_w f_{w0},\nonumber \\
a_2&=&\frac{f_{w0}}{\tau_c}+\frac{2}{\kappa}\frac{1-f_{w0}}{\tau_e}, \nonumber\\
a_3&=&\frac{\kappa}{2}\frac{1-f_{q0}}{\tau_c}+\frac{f_{q0}}{\tau_e},\nonumber\\
a_4&=&\frac{\kappa}{2}\frac{f_{w0}}{\tau_c}+\frac{1-f_{w0}}{\tau_e}
+2b_q f_{q0}+2g f_{S0}\nonumber\\
a_5&=&\frac{1}{\tau_\mathrm{ph}},\hspace{0.5in} \nonumber\\
a_6&=&2g\,f_{S0} \nonumber\\
a_7&=&1-f_{w0},
\label{eq:QDSSAparam}
\end{eqnarray}
in terms of various occupancies and timescales, already introduced in the 
description of Eqs.~(\ref{eq:fw})-(\ref{eq:fS}). 
The subscript 0 represents steady-state solutions.
By solving Eq.~(\ref{eq:QDSSA}), 
we obtain the response function for QD lasers, 
\begin{widetext}
\begin{eqnarray}
\left|\frac{R(\omega)}{R(0)}\right|_\mathrm{QD}
&=&
\left| \frac{a_1a_5a_6}{a_1a_5a_6-i\omega(a_1a_4-a_2a_3+a_5a_6)
-\omega^2(a_1+a_4)+i\omega^3} \right| \label{eq:QDraw}\\
&=&
\left|\frac{\omega_{r}^{2}}{(1-i\omega\tau_{c}')[\omega_{r}^{2}
-i \omega(c_2+c_3\frac{c_4/c_1}{1-i\omega\tau_{c}'})-
\omega^2(1+\frac{c_4/c_1}{1-i\omega\tau_{c}'}) ]}\right| \label{eq:QDexact}\\
&\approx&
\frac{\omega_{r}'^{2}}{(1+\omega^2\tau_{c}'^2)^{1/2}
[(\omega_{r}'^{2}-\omega^2)^2+\omega^2\gamma_\mathrm{QD}^2]^{1/2}\label{eq:QDappx}
},
\end{eqnarray}
\end{widetext}

where $\tau_c'=1/a_1$, $\omega_{r}^2=a_5a_6$, 
$c_1=\tau_c a_1$,
$c_2=a_4-(\kappa/2)a_2$,
$c_3=a_1-(2/ \kappa)a_3$, and
$c_4=(\kappa/2)\tau_c a_2$. 
When $\tau_c'\ll 1/\omega_r'$, 
we can approximate Eq.~(\ref{eq:QDexact}) as 
Eq.~(\ref{eq:QDappx}), 
where $\omega_{r}'^2=\omega_r^2/(1+c_4/c_1)$ and $\gamma_\mathrm{QD}=(c_2+c_3c_4/c_1)/(1+c_4/c_1)$, 
analogous to the same approximation in separate confinement heterostructure lasers.\cite{Nagarajan1992:JQE}
Then, bandwidth can be also easily obtained from Eq.~(\ref{eq:O3QD}).
However, since $\tau_c$ used in the mapping lies in a 
wider range, we employed a more general 
form of response function in Eq.~(\ref{eq:QDexact}) 
to find the bandwidth and study
the dynamic response of QD lasers. 

Within the parameter space used in this paper, 
several parameters from 
Eq.~(\ref{eq:QDSSAparam}) can be approximated as
\begin{eqnarray}
a_1&\approx&\frac{1-f_{q0}}{\tau_c}+2 b_w f_{w0},\nonumber\\
a_2&\approx&\frac{f_{w0}}{\tau_c}, \nonumber\\
a_3&\approx&\frac{\kappa}{2}\frac{1-f_{q0}}{\tau_c},\nonumber\\
a_4&\approx&\frac{\kappa}{2}\frac{f_{w0}}{\tau_e}+2b_q f_{q0}+2g f_{S0},  \nonumber\\
a_7&\approx&1.
\label{eq:QDSSAparamAPPROX} 
\end{eqnarray}

While in this work we have focused on quadratic recombination 
(quadratic in the carrier density), 
this consideration can be easily generalized.\cite{Gothgen2008:APL}  For linear recombination, 
the recombination time $\tau_r$  
in QW lasers is converted to a quadratic recombination rate $B$ 
such that the magnitude of the lasing threshold is preserved, 
\begin{eqnarray}
J_T&=&n_T/\tau_r=B n_T^2,\label{eq:B}
\end{eqnarray}
where $n_{T}=n_{\mathrm{tran}}+1/\Gamma g_0 \tau_\mathrm{ph}$ and the 
subscript $_T$ represents threshold. Using the above equality, 
one can convert $\tau_r$ to $B$, or vice versa.

Analogously to Eq.~(\ref{eq:B}), the recombination rates $b_q$ and $b_w$ that respectively 
arise in the WL and QD regions, can be calculated from the corresponding 
time constants $\tau_{rq}$ and $\tau_{rw}$ for an unchanged threshold. 
We first find $b_q$ as a function of $\tau_{rq}$ by assuming  
$\tau_{rw}=0$, and then $b_w$ is calculated from $\tau_{rw}\neq0$. Therefore,
the conversion equations in QD lasers are obtained that preserve the threshold: 
\begin{eqnarray}
b_q\, \tau_{rq}&=& \frac{2g\, \tau_\mathrm{ph}}{1+g\, \tau_\mathrm{ph}}, \nonumber\\
b_w\, \tau_{rw}&=&\frac{\tau_{rq}[2\tau_c(1+g\, \tau_\mathrm{ph})
+\kappa \tau_e(g\, \tau_\mathrm{ph}-1)]}{2\tau_c(g\, \tau_\mathrm{ph}+1)(\tau_e+\tau_{rq})}. 
\nonumber
\end{eqnarray}


\begin{thebibliography}{99}
\bibitem{negative} 
We mostly consider $P_J\ge0$ while the results for $P_J<0$ can be deduced easily.
\bibitem{const}
This is not essential but simplifies our analytical results.
\end{thebibliography}
\end{document}